\newcommand{\bra}[1]{\langle #1 |}
\newcommand{\ket}[1]{| #1 \rangle}

\documentclass[twocolumn,prb,showpacs,preprintnumbers,amsmath]{revtex4}

\usepackage{graphicx}

\begin{document}
\title{Entanglement and quantum phase transition in the asymmetric Hubbard chain: density-matrix renormalization group calculations}

\author{Wen-Ling Chan}
\author{Shi-Jian Gu}
 \email{sjgu@phy.cuhk.edu.hk}
 \homepage{http://www.phystar.net}
\affiliation{Department of Physics and Institute of Theoretical Physics, The Chinese University of Hong Kong, Hong Kong, China}

\date{\today}

\begin{abstract}
We study the ground state quantum phase transition by means of entanglement in the one-dimensional asymmetric Hubbard model with open boundary condition. The local entanglement between the middle two sites and the rest of the system, and the block entanglement between the left and right portions of the system, are calculated using the density-matrix renormalization group (DMRG) method. We find that the entanglement shows interesting scaling and singular behavior around the phase transition line.
\end{abstract}


\maketitle

\section{Introduction}

In condensed-matter physics, given a quantum system, the challenging problems are to find the ground state and to study the quantum phase transition \cite{ref:Sachdev}. For the models whose analytical solutions do not exist, the general approach is to characterize an order parameter and measure it from the ground state obtained numerically from a finite system, then do scaling studies and extrapolate the results to the thermodynamic limit. However in some cases this method is quite inefficient or even no definite conclusion can be drawn.

Recently it had been shown that entanglement \cite{ref:Einstein} may be an effective indicator of quantum phase transition in spin systems \cite{ref:Osterloh,ref:Vidal}. The concept was also applied to fermionic systems including the extended Hubbard model \cite{ref:Lin,ref:Gu,ref:Anfossi}, ionic Hubbard model \cite{ref:Legeza}, asymmetric Hubbard model \cite{ref:Gu2}, etc.\ \cite{ref:Larsson} Gu \textit{et al.}\ \cite{ref:Gu} studied the extended Hubbard model and found that the local entropy, that is, the entanglement between one site and the rest of the system, clearly indicates that phase transition occurs where the entropy is extremum or its derivative is singular. Legeza \textit{ et al.}\ \cite{ref:Legeza} showed that in some models the two-site entropy is a better indicator. These entropies can be readily obtained using the density-matrix renormalization group (DMRG) method \cite{ref:White,ref:Schollwock}.

In this paper we use entanglement to witness the phase transition of the one-dimensional asymmetric Hubbard model (AHM) with open boundary condition. In the AHM, there are two species of fermions, say, spin $\uparrow$ and spin $\downarrow$ particles, with different masses. The model is intensively studied recently as it may be used to describe some important physical properties in strongly correlated systems such as superconducting cuprates \cite{ref:Emery} and heavy fermionic systems \cite{ref:Varma}. It can be realized in experiments using cold fermionic atoms trapped in optical lattices \cite{ref:Liu,ref:Cazalilla,ref:Gu2}, where all model parameters can be tuned. It has also been studied theoretically \cite{ref:Gu2,ref:Fath,ref:Ueltschi,ref:Macedo,ref:Wang}. However, the complete phase diagram is still not very clear.

We consider equal number of both species of fermions. We numerically calculate the two-site entropy, as well as the block entropy \cite{ref:Vidal} which is defined as the entanglement between the left and right portions of the system. We propose that the entanglement shows interesting scaling behavior around the phase transition line.

The Hamiltonian of AHM reads
\begin{equation}
H = -\displaystyle\sum_{\langle i j \rangle} \displaystyle\sum_{\sigma=\uparrow,\downarrow} t_\sigma c_{i \sigma}^\dagger c_{j \sigma} + U \displaystyle\sum_i n_{i \uparrow} n_{i \downarrow}, \label{eq:H}
\end{equation}
where $t_\uparrow \geq t_\downarrow \geq 0$ are the hopping integrals for the light $\uparrow$ and heavy $\downarrow$ fermions respectively, $c_{i \sigma}^\dagger$ and $c_{i \sigma}$ are the creation and annihilation operator respectively, $U > 0$ is the on-site repulsive Coulomb interaction between the two species of fermions, and $n_{i \sigma} = c_{i \sigma}^\dagger c_{i \sigma}$ is the number operator. Hereafter we set $t_\uparrow=1$. The AHM reduces to the Hubbard model (HM) \cite{ref:Hubbard,ref:Lieb,ref:Essler} for $t_\uparrow = t_\downarrow = 1$ and to the Falicov-Kimball model (FKM) \cite{ref:Falicov,ref:Brandt,ref:Lemberger,ref:Kennedy,ref:Freericks} for $t_\downarrow=0$. We only consider the cases with $N_\uparrow = N_\downarrow$. The filling density is defined as $n = (N_\uparrow + N_\downarrow)/L$, where $L$ is the length of the chain, or the number of sites. Half-filling ($n = 1$) is achieved when $N_\uparrow = N_\downarrow = L/2$.

It is well-known that the HM and FKM belong to different universality classes \cite{ref:Fath}. A phase transition should occur somewhere on the $U$-$t_\downarrow$ plane \cite{ref:Fath,ref:Ueltschi,ref:Cazalilla}. To understand the phases, first we look at the perturbation expansion in the large-$U$ limit. When $t_\uparrow, t_\downarrow \ll U$, the hopping term in Hamiltonian~(\ref{eq:H}) can be regarded as a small perturbation. For the HM, provided that $n \leq 1$, the expansion leads effectively to the t-J model \cite{ref:Lin2}. The method can be generalized to the AHM \cite{ref:Fath}. Since the calculation is straightforward but lengthy, we present the final result only. The expansion leads effectively to the anisotropic Heisenberg model with hopping. Explicitly the effective Hamiltonian reads
\begin{equation}
\begin{split}
& H_\text{eff} = -\displaystyle\sum_{\langle i j \rangle} \displaystyle\sum_{\sigma=\uparrow,\downarrow} t_\sigma \tilde{c}_{i \sigma}^\dagger \tilde{c}_{j \sigma} + \\
& + \frac{t_\uparrow t_\downarrow}{U} \displaystyle\sum_{\langle i j \rangle} \left[ \sigma_i^x \sigma_j^x + \sigma_i^y \sigma_j^y + \Delta ( \sigma_i^z \sigma_j^z - 1 ) \right] + O \left( \frac{t_\sigma^4}{U^3} \right), \label{eq:Heff}
\end{split}
\end{equation}
where
\begin{eqnarray}
\tilde{c}_{i \sigma}^\dagger = ( 1 - n_{i \sigma} ) c_{i \sigma}^\dagger, \\
\Delta = \frac{t_\uparrow^2 + t_\downarrow^2}{2 t_\uparrow t_\downarrow} \geq 1.
\end{eqnarray}
At $n = 1$, $\tilde{c}_{i \sigma}^\dagger$ and $\tilde{c}_{j \sigma}$ can be approximated as zero in the large-$U$ limit, hence the hopping term in the effective Hamiltonian vanishes and the effective model becomes the anisotropic Heisenberg model (XXZ model). Therefore, the system behaves very differently at $n = 1$ and $n < 1$. The case $n > 1$ can be treated by considering the particle-hole symmetry \cite{ref:Kennedy2} of the AHM. It is found that the energy spectrum is invariant (except for a global shift by a constant) under the exchange of the numbers of particles and holes, hence the physical properties of the system at $n > 1$ are just the same as that at $n < 1$.

Away from half-filling (either $n < 1$ or $n > 1$), the system possesses the density wave (DW) phase and phase separation (PS) phase \cite{ref:Emery2,ref:Lemberger}; at half-filling ($n=1$), the system possesses effectively the XY phase and Ising phase.

The paper is arranged as follows. In section \ref{sec:entnaglement}, we show how entanglement is measured and implemented in the DMRG algorithm. In section \ref{sec:results}, we give the numerical results and discussions for the away-from-half-filling and half-filling cases respectively. Finally a summary is given in section \ref{sec:summary}.

\section{\label{sec:entnaglement}Measurement of entanglement}

We are interested in the entanglement between a local block, which is composed of one or more sites, and the rest of the system. For the AHM, the local state on each site has four possible configurations: $\ket{0},\ket{\uparrow},\ket{\downarrow}$ and $\ket{\uparrow\downarrow}$. The Hilbert space associated with the system with $L$ sites is spanned by $4^L$ basis vectors. Suppose we have obtained the ground state $\ket{\Psi}$, the reduced density matrix of the local block with $l$ sites is
\begin{equation}
\rho_l = \text{tr}_{L-l}\ket{\Psi}\bra{\Psi}. \label{eq:rho}
\end{equation}
This matrix can be expressed in block diagonal form due to the fact that the numbers of $\uparrow$ and $\downarrow$ fermions are respectively conserved:
\begin{equation}
\rho_l = \text{diag} \{ (0,0), (1,0), (0,1), (1,1), ... , (l,l) \},  \label{eq:rho_diag}
\end{equation}
where $(m,n)$ means a block whose bases contain $m$ $\uparrow$ and $n$ $\downarrow$ fermions. The von Neumann entropy
\begin{equation}
S_l = -\text{tr} \left[ \rho_l \log_2 (\rho_l) \right]  \label{eq:Ev}
\end{equation}
measures the entanglement between the $l$ sites and the rest $L-l$ sites of the system. In general, the more evenly distributed the eigenvalues of $\rho_l$, the higher the entropy is. The degree of freedom (and hence number of bases) within a block of length $l$ is $4^l$, therefore there are $4^l$ eigenvalues. The entropy $S_l$ is maximum when all eigenvalues are equal ($=4^{-l}$), and so $S_{l,\max} = -\log_2 4^{-l} = 2 l$.

We use the density-matrix renormalization group (DMRG) method \cite{ref:White,ref:Schollwock} to calculate the ground state properties of the system. This method is efficient and accurate for one-dimensional lattice models. It involves iterative diagonalization of a Hamiltonian in a approximated, size-limited Hilbert space to obtain the target state (usually the ground state). The approximated Hilbert space is constructed from an appropriate number of eigenvectors of the reduced density matrix of a part of the system.

We are interested in two quantities: the two-site entropy and the block entropy. Conceptually they are the same kind of measurement. Both of them are the von Neumann entropy of the reduced density matrix of a part of the chain. The two-site entropy, denoted as $S_2$, is defined as the entanglement between the middle two consecutive sites with the rest of the system. We choose the middle sites so as to minimize any boundary effects. Note that we consider the chain with even number of sites only. The measurement can be easily implemented in the DMRG method \cite{ref:Legeza}. During a ``sweep'' in the finite lattice algorithm, when the two free sites ``move'' to the middle, we obtain the reduced density matrix of them and compute the von Neumann entropy using Eq.~(\ref{eq:Ev}).

On the other hand, the block entropy, denoted as $S(l)$, is defined as the entanglement between the left block consisting of $l$ sites and the remaining right block. Studies of this quantity have established a bridge between the quantum information theory and conformal field theory \cite{ref:Vidal,ref:Korepin,ref:Calabrese}. In the DMRG algorithm the left and right blocks are renormalized in each step. The well-known fact that the entropy does not change upon renormalization group transformation enables us to obtain the required block entropy.

In the DMRG method we apply the dynamical block selection approach \cite{ref:Legeza2} plus information loss control \cite{ref:Legeza3} to increase the efficiency and accuracy. In the following computations we set the information loss $\chi < 10^{-8}$ and the minimum and maximum number of DMRG states to be 100 and 250 respectively. Also, we apply the seed vector construction routine \cite{ref:White2} to further improve the efficiency.

\section{\label{sec:results}Numerical results}

\subsection{Away from half-filling}

In this section we examine the system where the total number of particles does not equal $L$. In particular, we study 1/4-filling, that is $N_\uparrow = N_\downarrow = L/4$ and $n=1/2$. Before presenting the results of the entanglement measurements, we would like to clarify the phases first.

\begin{figure}
\includegraphics[width=9.0cm]{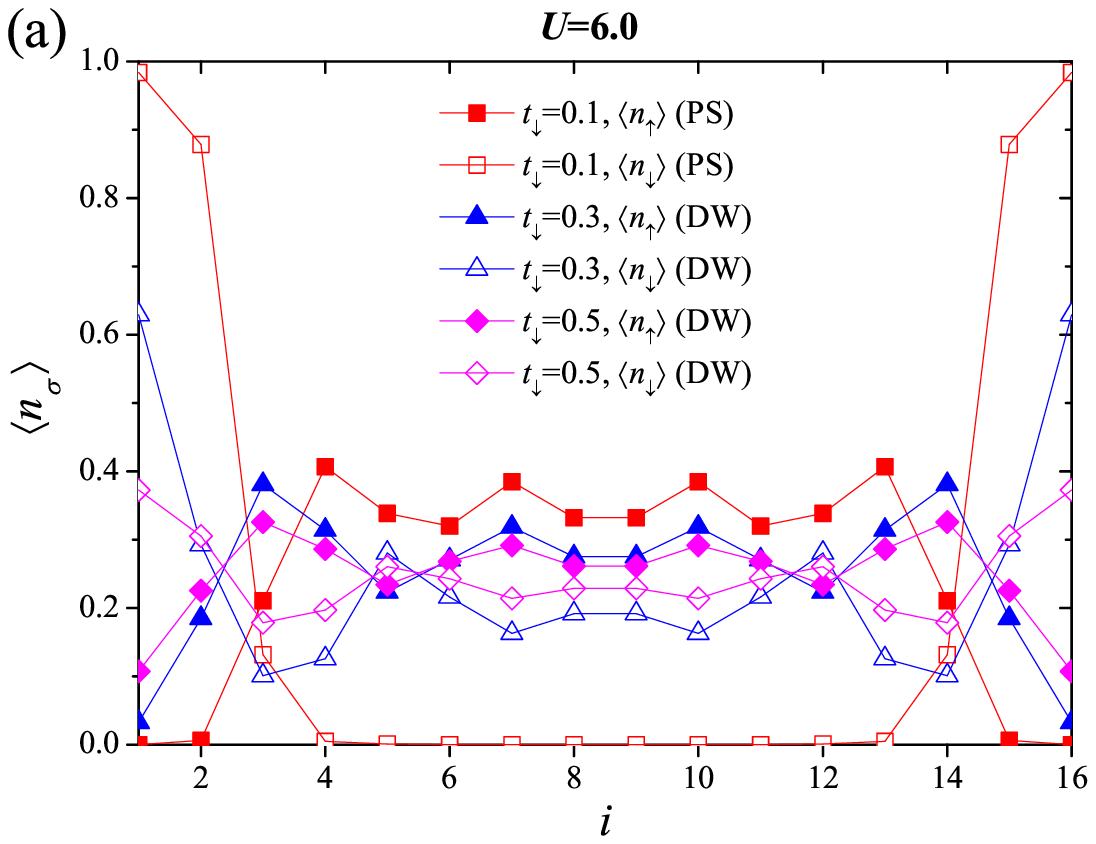} \\
\includegraphics[width=9.0cm]{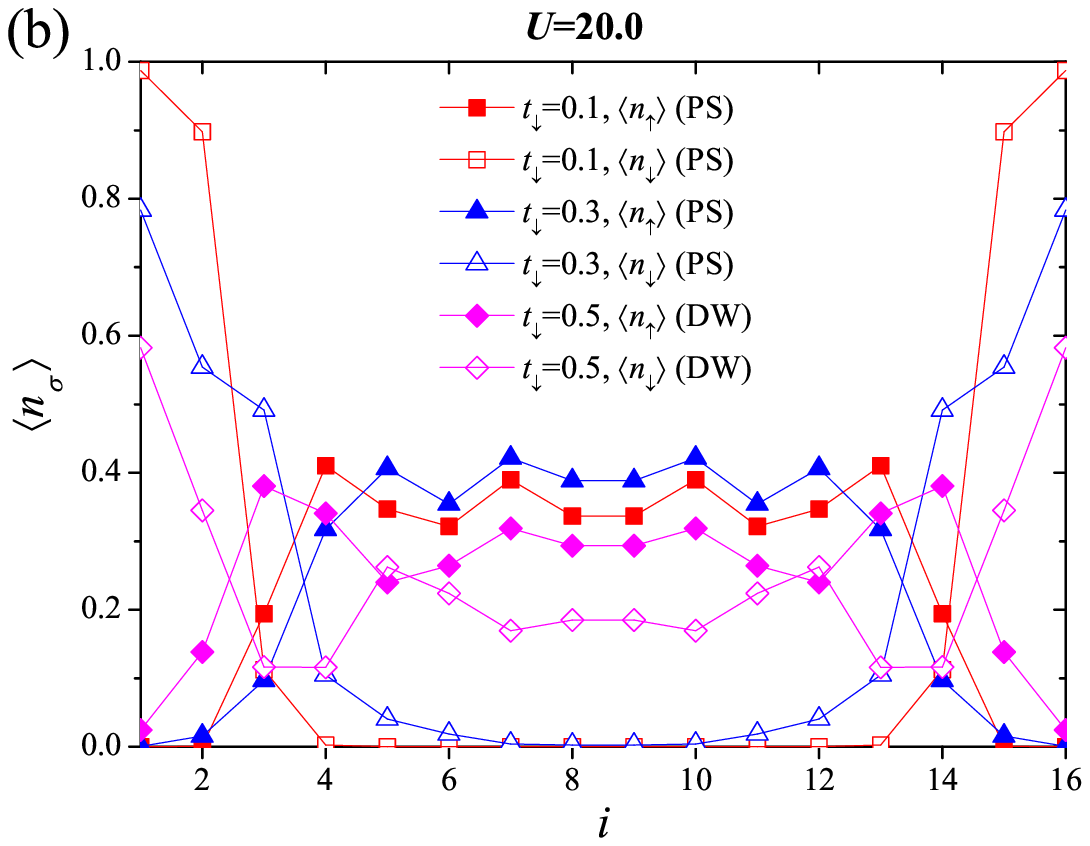}
\caption{\label{fig:wdmrg_n}(Color online) The number densities of $\uparrow$ and $\downarrow$ fermions at site $i$ for a chain of length $L=16$ at $1/4$-filling. (a) $U=6.0$, (b) $U=20.0$.}
\end{figure}

When $t_\uparrow$ is much larger than $t_\downarrow$ in magnitude, the hopping of the light $\uparrow$ fermions becomes much more helpful to lower the system's energy than that of the heavy $\downarrow$ fermions. The appropriate configuration is that the light fermions spread around in a large pool of free sites while the heavy fermions congregate together. The two species of fermions separate, hence the name phase separation (PS) phase to describe the system. For open boundary condition, as in our case, the light fermions tend not to stay at the ends of the chain as it is too costly to sacrifice some freedoms of hopping, hence they are the heavy ones which fill the ends. The following shows a typical dominant configuration in this phase (24 sites, $1/4$-filling):
\begin{equation}
\ket{\downarrow \downarrow \downarrow 0 \uparrow 0 0 \uparrow 0 \uparrow 0 0 \uparrow 0 0 \uparrow 0 0 \uparrow 0 0 \downarrow \downarrow \downarrow}
\end{equation}
On the other hand, when $t_\uparrow \approx t_\downarrow$, the hopping of the two species of fermions are equally important. It turns out that both of them distribute uniformly on the whole chain and the system is in the so called density wave (DW) phase. A typical dominant configuration is:
\begin{equation}
\ket{\downarrow 0 \uparrow 0 \downarrow \uparrow 0 \downarrow 0 0 \uparrow 0 \downarrow 0 \uparrow 0 \downarrow 0 0 \uparrow \downarrow 0 \uparrow 0}
\end{equation}
Our conjecture can be verified by measuring the local densities of the $\uparrow$ and $\downarrow$ fermions. Fig.~\ref{fig:wdmrg_n} shows the results for a chain of 16 sites with 4 $\uparrow$ and 4 $\downarrow$ fermions. Clearly, for small $t_\downarrow$, the density of the heavy $\downarrow$ fermions is much higher at the two ends of the chain (PS phase), while for large $t_\downarrow$, the density distribution is more even (DW phase).

\begin{figure}
\includegraphics[width=9.0cm]{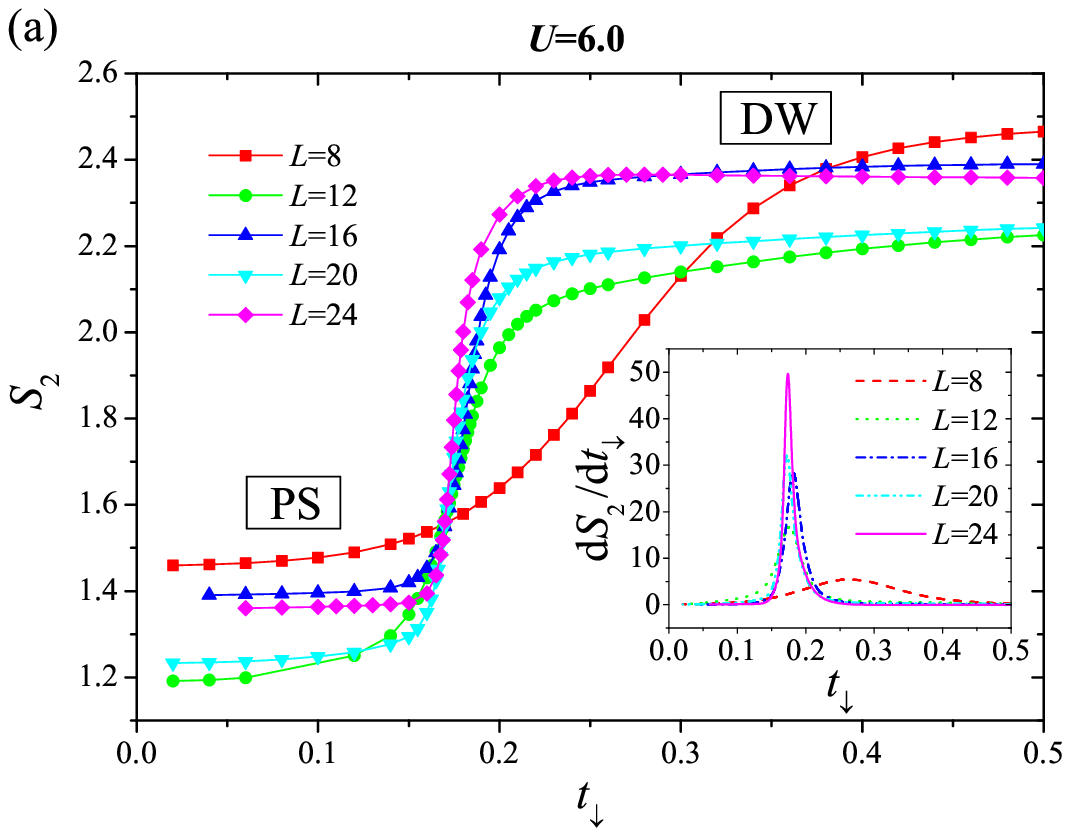} \\
\includegraphics[width=9.0cm]{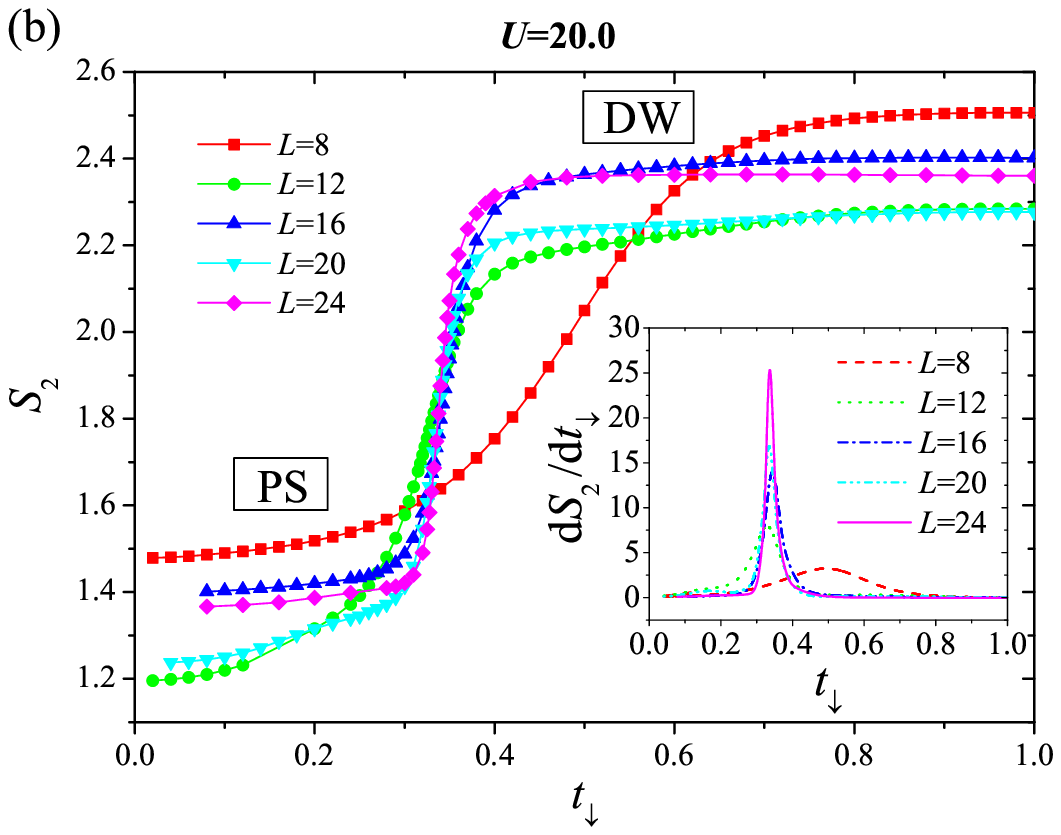}
\caption{\label{fig:wdmrg}(Color online) The two-site entropy $S_2$ against $t_\downarrow$ for different chain length $L$ at $1/4$-filling. (a) $U=6.0$, (b) $U=20.0$. The first derivatives are shown in the insets.}
\end{figure}

Then we compute the two-site entropy $S_2$ for different values of $t_\downarrow$ and $U$ using the DMRG method. Fig.~\ref{fig:wdmrg}(a) and (b) show the results for $U=6.0$ and $U=20.0$ respectively. It is obvious that, for any given $L$, two plateaus appear. For small $t_\downarrow$ (PS phase) $S_2$ is lower, while for large $t_\downarrow$ it is higher. The cliff connecting the two plateaus becomes steeper and steeper as $L$ increases. This property is clearly revealed when we plot the first derivatives of the curves in the insets. For any fixed $L$, a peak appears between the two phases. The peak is sharper for a longer chain. It may be expected that in the thermodynamic limit the peak goes to infinity, indicating that this phase transition is a typical Landau one.

Comparing the $U=6.0$ and $U=20.0$ cases, the critical $t_\downarrow$ is larger for the larger $U$. This result is consistent with Ref.~\onlinecite{ref:Gu2}. For the case $U=6.0$, all curves seem to pass through the same point at $t_\downarrow=0.170$. We check that, however, this does not happen for other values of $U$. For instance, for $U=1.0$ (which is not shown here), some curves do not even cross in the critical regime. Therefore the special crossing in the case $U=6.0$ should be merely a coincidence.

It is not difficult to understand the phenomenon that the two-site entropy for the PS phase is lower than that of the DW phase. According to Eq.~(\ref{eq:rho_diag}), the two-site reduced density matrix $\rho_2$ consists of 9 blocks. In the PS phase, the heavy $\downarrow$ fermions congregate at the two ends while the light $\uparrow$ fermions distribute in the middle. Only the blocks (0,0), (1,0) and (2,0) contain significant values, hence $S_2$ is low. However in the DW phase, the blocks (0,0), (1,0), (0,1) and (1,1) contain significant values and so $S_2$ is higher.

\begin{figure}
\includegraphics[width=9.0cm]{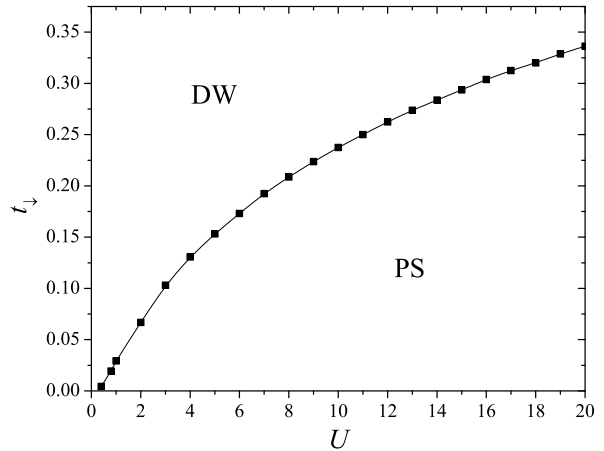}
\caption{\label{fig:phaseline} Ground state phase diagram of AHM at $1/4$-filling in the thermodynamic limit, deduced by the singular behavior of the two-site entropy. The error bars are smaller than the size of the symbols.}
\end{figure}

The above study shows that the two-site entropy $S_2$ is a good indicator of phase transition. From the insets of Fig.~\ref{fig:wdmrg}, it is obvious that at a given $U$ the critical $t_\downarrow$ converges when $L$ increases. The results can be extrapolated to the thermodynamic limit. We repeat the same analysis for different $U$ and the phase transition line in the $U$-$t_\downarrow$ plane can be obtained, as presented in Fig.~\ref{fig:phaseline} with error bars smaller than the size of the symbols.

\begin{figure}
\includegraphics[width=9.0cm]{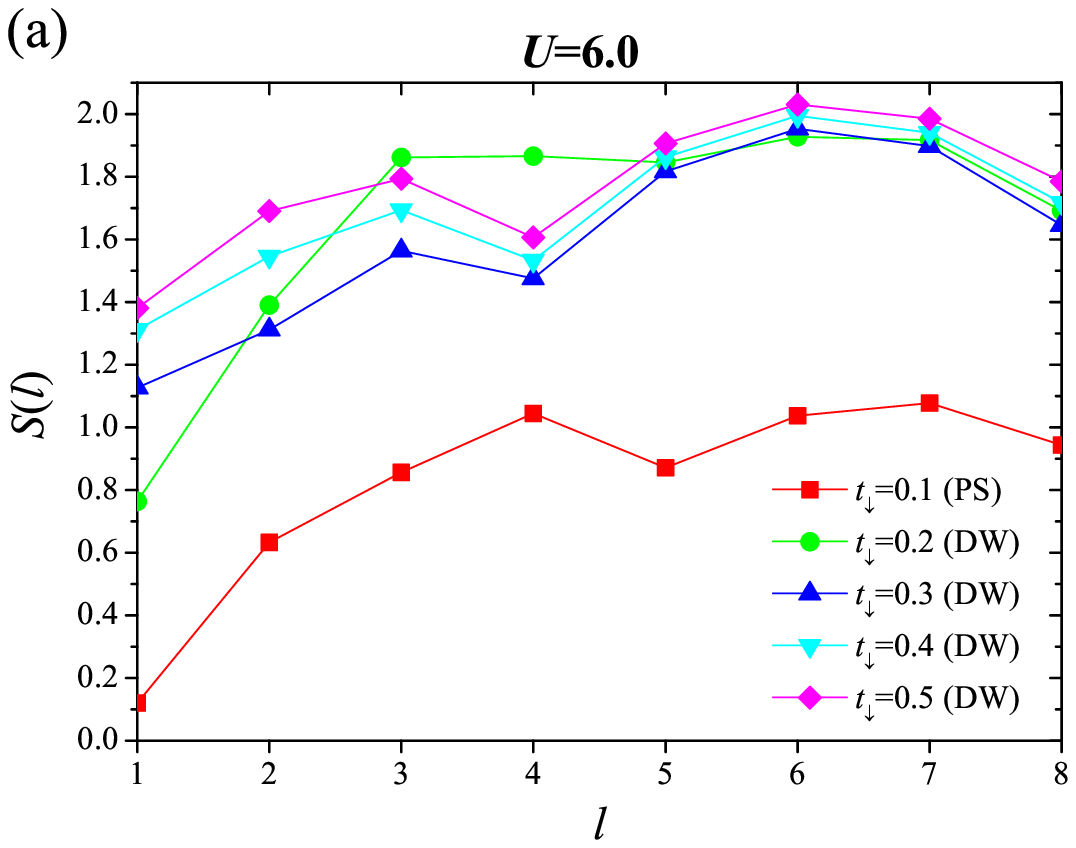} \\
\includegraphics[width=9.0cm]{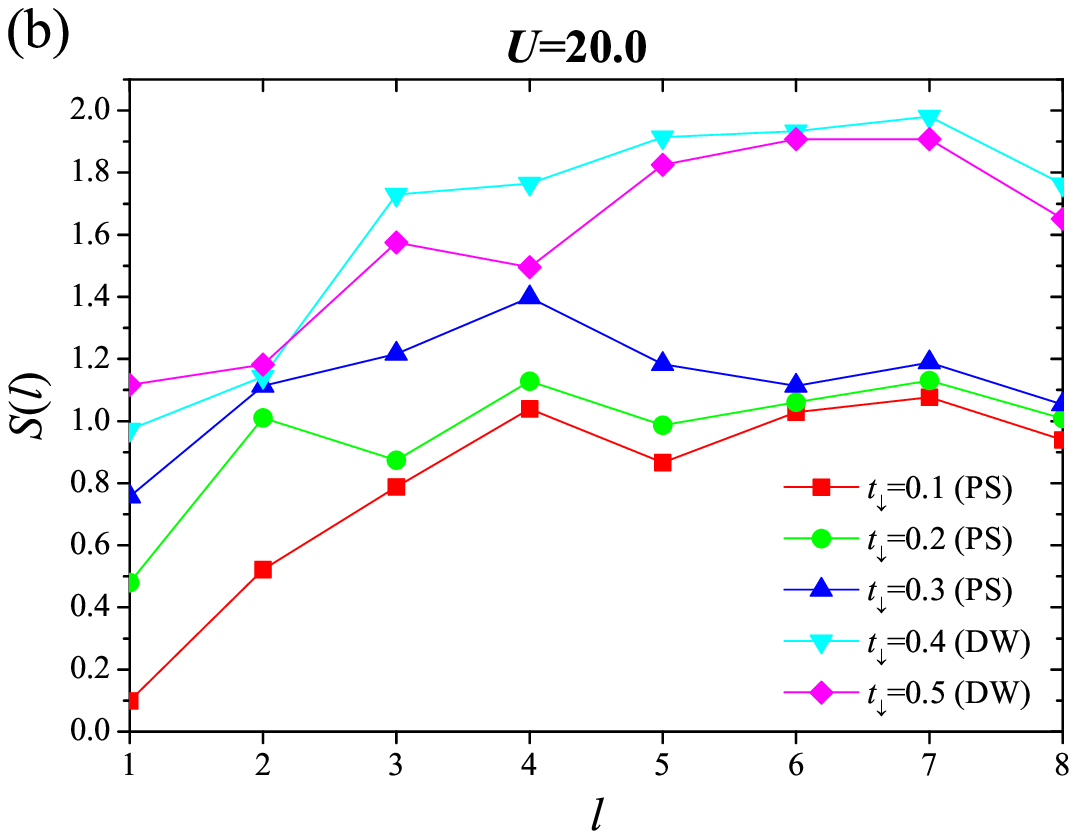}
\caption{\label{fig:wdmrg_BEv}(Color online) The block entropy $S(l)$ for different $t_\downarrow$ at 1/4-filling in a 16-site chain. (a) $U=6.0$, (b) $U=20.0$.}
\end{figure}

Next, we examine the block entropy of the same system. Fig.~\ref{fig:wdmrg_BEv} shows the results for a 16-site chain. Due to the symmetry $S(l)=S(L-l)$, it is sufficient to show $S(l)$ for $l=1,...,L/2$. First we note that in general $S(l)$ increases with $l$. It is not surprising because when the block becomes longer, there are more finite elements in the reduced density matrix $\rho_l$ and hence higher $S(l)$. On the other hand, the block entropy in the DW phase is always higher than that in the PS phase. It is because, in the PS phase, only the matrix elements (in $\rho_l$) related to the bases where the $\downarrow$ fermions congregate to the left are significant, however, in the DW phase, much more elements are significant.

For small $l$, $S(l)$ fluctuates due to boundary effects, whereas it becomes more stable for long $l$. Moreover, the results show that $S(l)$ for different $t_\downarrow$ converge for long $l$. This indicates that the entropy of a longer block is a better indicator for phase transition. This conclusion is consistent with that in Ref.~\onlinecite{ref:Legeza}. It is not surprising as the block entropy of a longer block includes more correlation functions. It is a pity that the convergence in the DMRG algorithm is poor for the AHM away from half-filling. Results obtained from longer chains are not quite reliable.

\begin{figure}
\includegraphics[width=8.0cm]{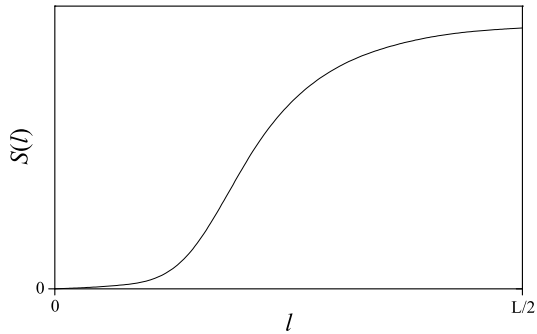}
\caption{\label{fig:wdmrg_PS} Proposed schematic sketch of the block entropy $S(l)$ against block length $l$ at the ground state in the PS phase of a very long chain away from half-filling.}
\end{figure}

For a very long chain we suggest that $S(l)$ should behave like that schematically sketched in Fig.~\ref{fig:wdmrg_PS}. The rise of $S(l)$ suppresses at the beginning as there is only one significant configuration (in which all sites contain the $\downarrow$ fermions) and hence $S(l)$ is close to zero. As the block grows and reaches the pool of $\uparrow$ fermions, $S(l)$ rises rapidly. For long $l$, the increase of the number of possible configurations is restricted due to the fact that the amount of the $\uparrow$ fermions is limited, hence $S(l)$ rises more slowly and soon saturates.

\subsection{Half-filling}

The physics of AHM at half-filling is quite different from that away from half-filling. In the large-$U$ limit, the system at half-filling possesses two phases: the Hubbard phase (effectively the XY phase) for large $t_\downarrow$ and the FK phase (effectively the Ising phase) for small $t_\downarrow$. The transition is known to be Kosterlitz-Thouless like \cite{ref:Yang,ref:Kosterlitz}. The ground state phase diagram on the $U$-$t_\downarrow$ plane, unlike that in the previous case of away from half-filling, is more subtle and hard to be completed. F\'{a}th \textit{et al.}\ \cite{ref:Fath} studied the perturbation expansion and found that in the large-$U$ limit the phase transition line is right on the Hubbard line $t_\downarrow = 1$. For small $U$ they computed the spin gap and the magnetic order parameter, however no solid conclusion could be drawn. In this section we give a try using entanglement.

\begin{figure}
\includegraphics[width=9.0cm]{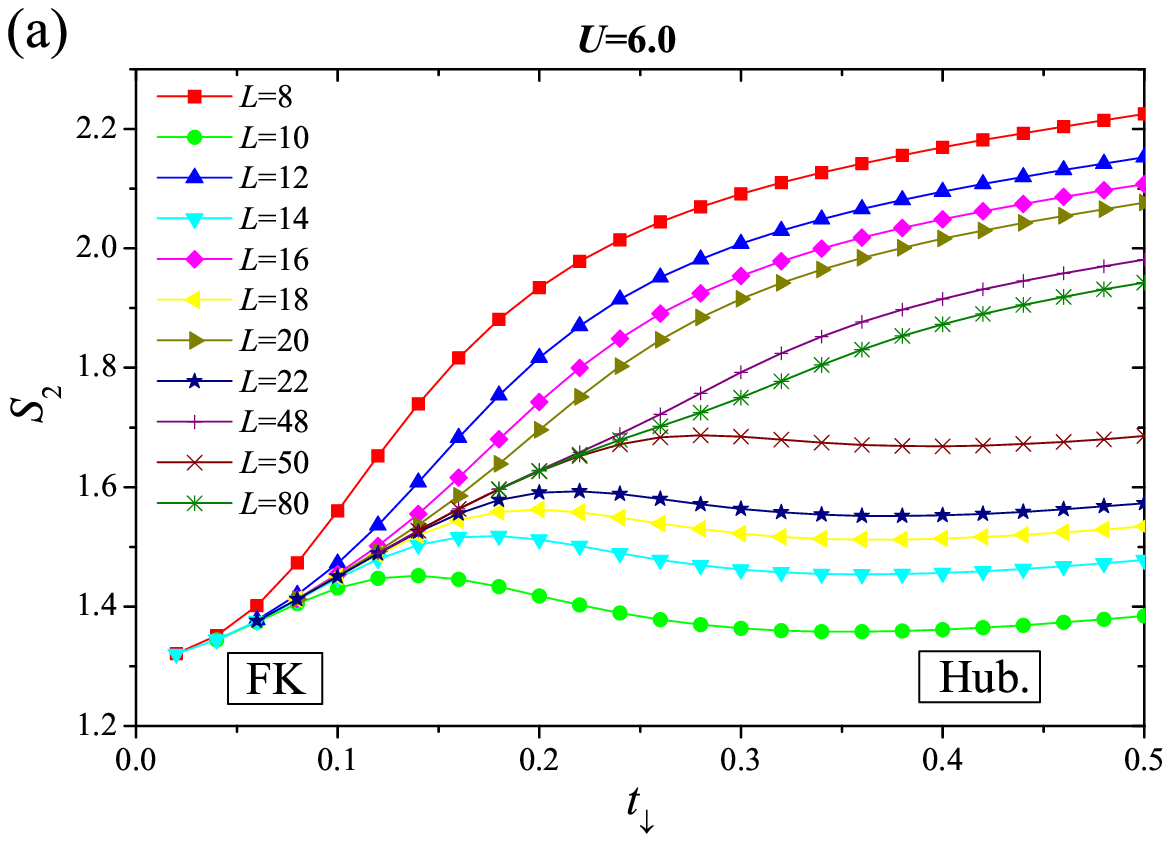}
\includegraphics[width=9.0cm]{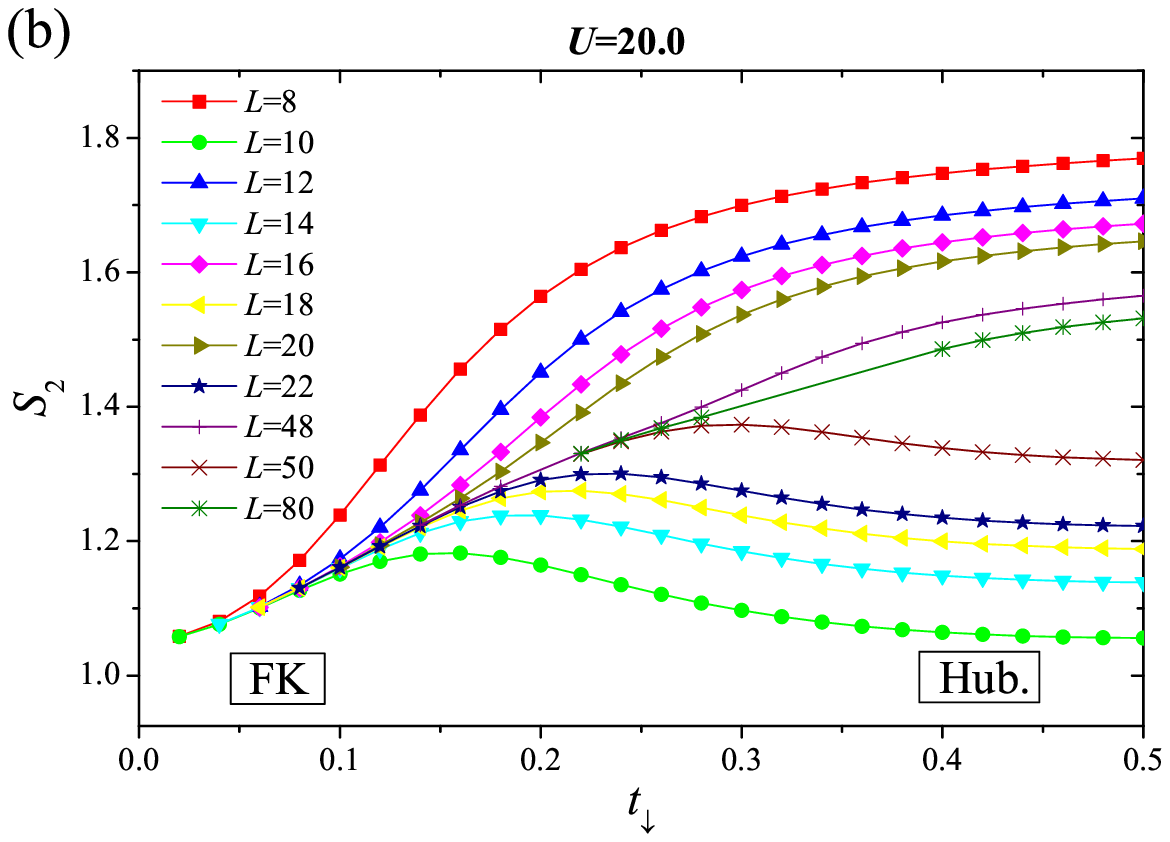}
\caption{\label{fig:dmrg}(Color online) The two-site entropy $S_2$ against $t_\downarrow$ for different chain length $L$ at half-filling. (a) $U=6.0$, (b) $U=20.0$.}
\end{figure}

First we compute the two-site entropy against $t_\downarrow$ for different $L$. The results are shown in Fig.~\ref{fig:dmrg}. There are roughly two regions. For small $t_\downarrow$, $S_2$ presents algebraic scaling with $t_\downarrow$ (FK phase), while for large $t_\downarrow$, $S_2$ scales linearly with $t_\downarrow$ (Hubbard phase). The curves do not show discontinuity for higher derivatives even for long $L$. This is a property of the Kosterlitz-Thouless like transitions \cite{ref:Yang,ref:Kosterlitz}.

\begin{figure}
\includegraphics[width=9.0cm]{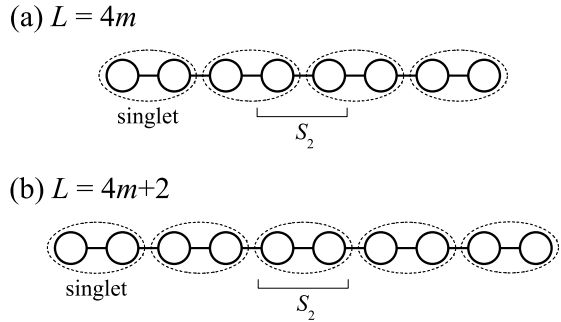}
\caption{\label{fig:singlet}Schematic diagrams showing the configurations of singlets. (a) For the family $L=4m$ (here $L=8$), the middle two sites are little correlated, hence higher the entropy $S_2$. (b) For the family $L=4m+2$ (here $L=10$), the middle two sites are strongly correlated, hence lower $S_2$.}
\end{figure}

The finite-size effects give rise to the formation of two families of curves. One family is $L=4m$ ($m=1,2,...$) which has higher $S_2$, while the other one is $L=4m+2$ which has lower $S_2$. The explanation is as follows. On the chain, the first site tends to form a singlet $(\ket{\uparrow \downarrow}-\ket{\downarrow \uparrow})/\sqrt{2}$ with the second site so as to lower the energy of the system. The third site tends to form a singlet with the forth one, and so on. Fig.~(\ref{fig:singlet}) schematically presents this phenomenon. If $L=4m$, the middle two sites are from different singlets and hence little correlated. The spin freedom is higher, the uncertainty is more and therefore the entropy $S_2$ is higher. On the other hand, if $L=4m+2$, the middle two sites tend to form a singlet and hence strongly correlated. The spin freedom is lower, the uncertainty is less and therefore $S_2$ is lower. We have verified this by measuring two-site entropies on different parts of the chain. When $L$ increases, the two families tend to merge together. It is expected that in the thermodynamic limit, they should be no longer distinct since all finite-size effects vanish.

Besides, the curves converge to a single point for $t_\downarrow \rightarrow 0$. In the $U \rightarrow \infty$ limit, the point is expected to be at $S_2=1.0$. This can be explained as follows. In the limits $U \rightarrow \infty$ and $t_\downarrow \rightarrow 0$, the effective Hamiltonian reduces to the Ising model. The ground state is doubly-degenerate and the two configurations are $\ket{\uparrow \downarrow \uparrow \downarrow ... \uparrow \downarrow}$ and $\ket{\downarrow \uparrow \downarrow \uparrow ... \downarrow \uparrow}$. For a finite system the ground state is the symmetric combination of the two, hence $S_2=1.0$ for any even $L$.

\begin{figure}
\includegraphics[width=9.0cm]{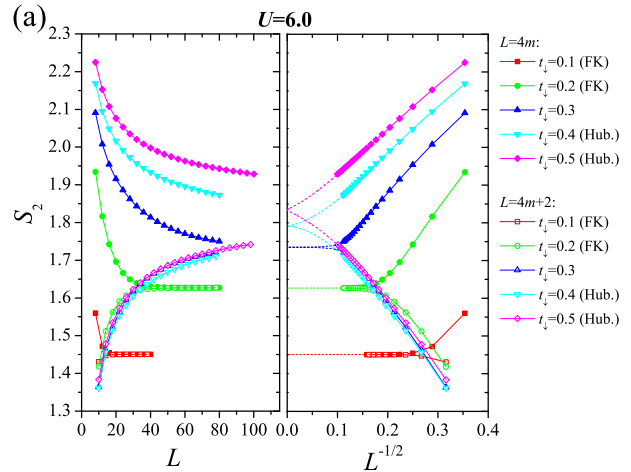}
\includegraphics[width=9.0cm]{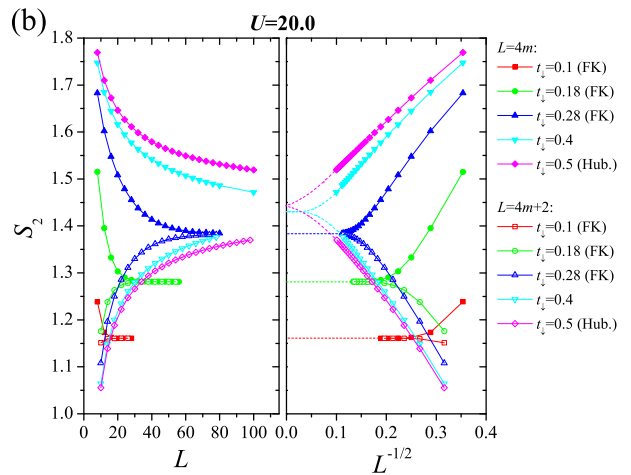}
\caption{\label{fig:dmrg_s}(Color online) The two-site entropy $S_2$ versus chain length $L$ or $1/\sqrt{L}$ for different $t_\downarrow$ at half-filling. (a) $U=6.0$, (b) $U=20.0$. Determinations of phase are suggested in the brackets.}
\end{figure}

The two-site entropy also presents different scaling behavior with $L$ in the two phases. Fig.~\ref{fig:dmrg_s} presents $S_2$ versus $L$ and $1/\sqrt{L}$. The two families of curves ($L=4m$ and $L=4m+2$) are indicated with distinct lines. From the lines' shape we suggest to which phase each line belongs. Some lines belong to the critical regime between the two phases. In the FK phase, $S_2$ goes like $A(L \rightarrow \infty)+B\exp(-L/\xi)$ where $A$, $B$ and $\xi$ are parameters that may depend on $t_\downarrow$, $U$ and $L$. $B$ is positive for $L=4m$ and negative for $L=4m+2$. For smaller $t_\downarrow$, $S_2$ converges more quickly. The physical interpretation is that the state is more ordered. In the Hubbard phase, $S_2$ goes like $C(L \rightarrow \infty)+D/\sqrt{L}$ where $C$ and $D$ may also depend on $t_\downarrow$, $U$ and $L$. $D$ is positive(negative) for $L=4m$($L=4m+2$).

We expect the two families of curves for a certain $t_\downarrow$ converge to the same value of $S_2$ in the thermodynamic limit, as the value should be independent of the parity and boundary condition of the chain. In the figures we extrapolate the curves and deduce the value in the limit.

\begin{figure}
\includegraphics[width=9.0cm]{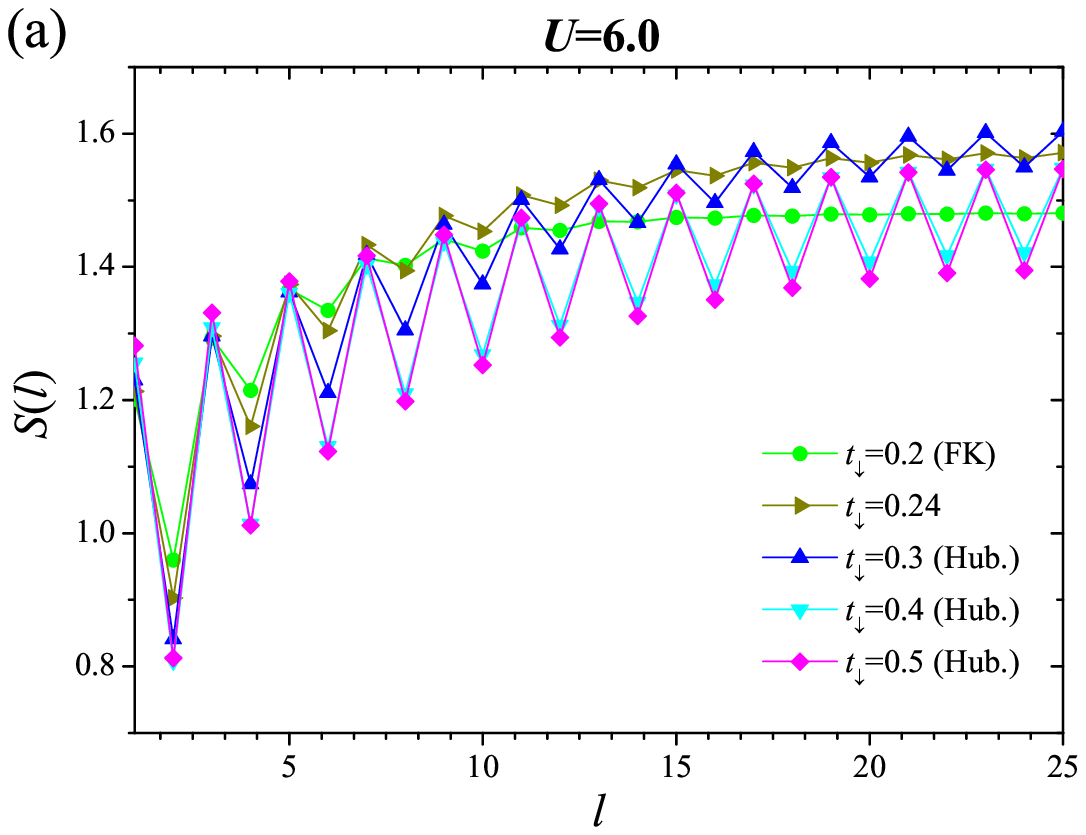}
\includegraphics[width=9.0cm]{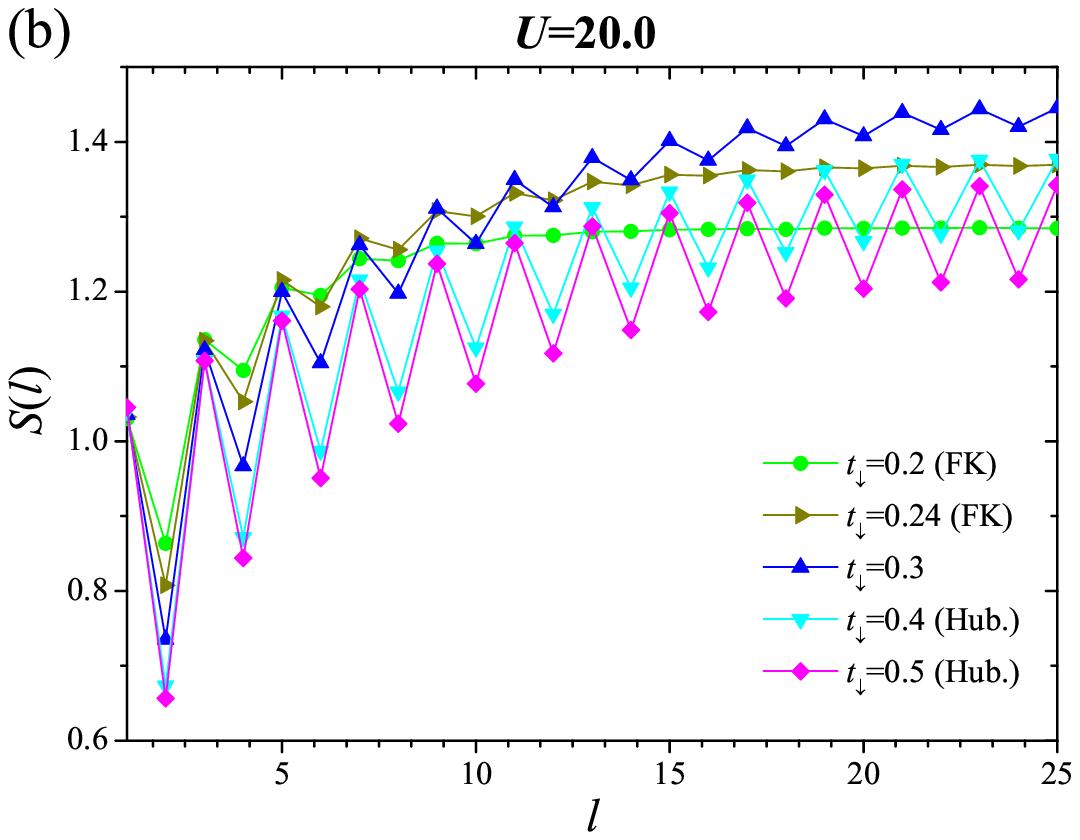}
\caption{\label{fig:dmrg_BEv}(Color online) The block entropy $S(l)$ for different $t_\downarrow$ in a half-filling 50-site chain. (a) $U=6.0$, (b) $U=20.0$. Determinations of phase are suggested in the brackets.}
\end{figure}

Second, we compute the block entropy. Fig.~\ref{fig:dmrg_BEv} shows the results for a 50-site chain. Again, due to the reflection symmetry $S(l)=S(L-l)$, it is sufficient to show $S(l)$ for $l=1,...,L/2$. In general $S(l)$ increases with $l$, for the same reason that is explained in the previous section for the case of away from half-filling: when the block grows, the size of the reduced density matrix $\rho_l$ increases and the number of finite elements in it increases, so $S(l)$ rises. On the other hand, the block entropy fluctuates with respect to the parity of $l$. For even $l$, the block is composed of a neatly arranged series of two-site singlets and hence $S(l)$ is lower. The fluctuation gets less for long $l$ as the inner sites have less tendency to form singlets.

From the lines' trend we suggest to which phase each line belongs. Some lines belong to the critical regime between the two phases. The behavior of $S(l)$ is obviously different in the two phases. In the Hubbard phase, for either parity of $l$, $S(l)$ rises algebraically, whereas in the FK phase, for either parity of $l$, $S(l)$ rises quickly and saturates because the state is more ordered. In the $U \rightarrow \infty$ and $t_\downarrow \rightarrow 0$ limits it should be expected that in the FK phase $S(l) \rightarrow 1.0$ for all $l$ because the ground state is the symmetric combination of the two definite Ising configurations: $\ket{\uparrow \downarrow \uparrow \downarrow ...}$ and $\ket{\downarrow \uparrow \downarrow \uparrow ...}$.

\begin{figure}
\includegraphics[width=9.0cm]{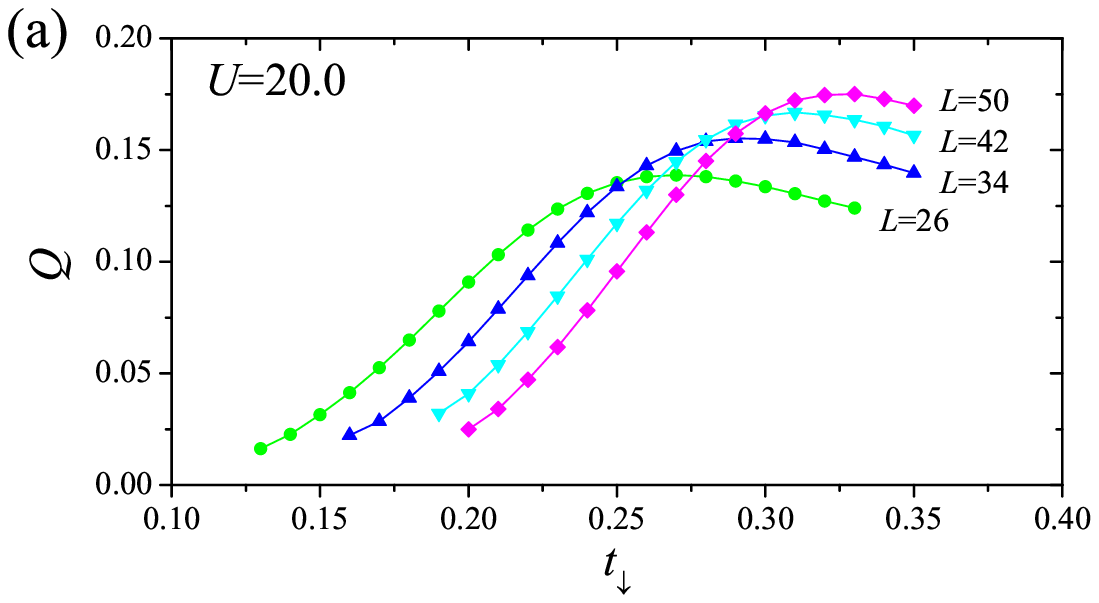}
\includegraphics[width=9.0cm]{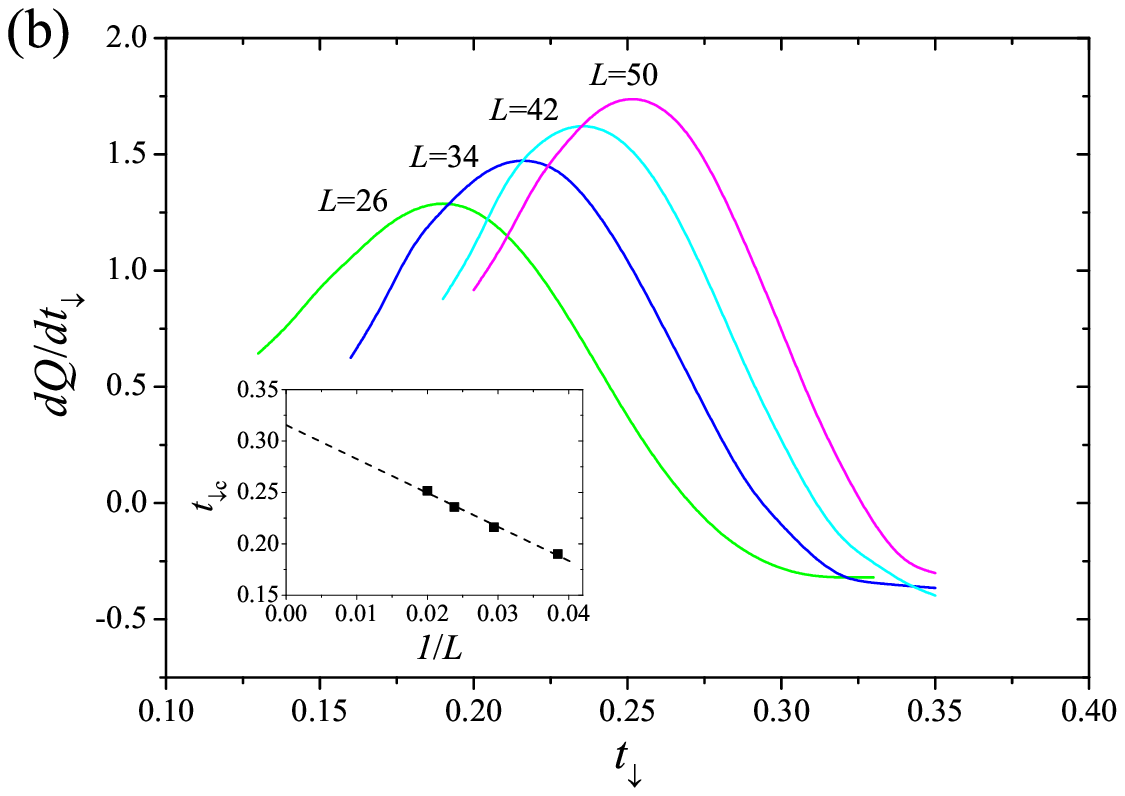}
\caption{\label{fig:dmrg_BEv_Q}(Color online) (a) The parameter $Q$, as calculated from Eq.~(\ref{eq:Q}), against $t_\downarrow$ for different chain length $L$ at half-filling, fixed $U=20.0$. (b) The first derivatives of the curves in (a). For each line the peak locates the critical point $t_{\downarrow c}$. Based on the scaling with $1/L$, the $t_{\downarrow c}$ in the thermodynamic limit can be obtained, as shown in the inset.}
\end{figure}

Vidal \textit{et al.}\ \cite{ref:Vidal} suggested that, due to conformal field theory,
\begin{equation}
S(l) \approx P + Q \ln l    \label{eq:ln}
\end{equation}
for long $l$, where $P$ and $Q$ are parameters that may depend on $t_\downarrow$, $U$ and $L$. In the ``non-critical regime'' (FK phase here) $Q \approx 0$, while in the ``critical regime'' (Hubbard phase here) $Q$ is proportional to the ``central charge'' of the conformal field theory and is finite. Therefore, the parameter $Q$ can be regarded as an indicator for the phase transition. For finite systems where boundary effects are noticeable, Eq.~(\ref{eq:ln}) is valid only for $l$ in the vicinity of $L/4$. Hence, we take $Q$ as the value of the slope of the $S(l)$-$\ln l$ graph at $l \approx L/4$. Only the data points at odd $l$ are used so that the fluctuation of the graph due to the open boundary effect does not affect our result. We pick three data points to measure the slope. Thus, $Q$ is computed as the weighted average of the slope between the first and second point and the slope of the second and third point. Explicitly,
\begin{eqnarray}
l_0 = [L/4], \nonumber \\
l_\pm = l_0 \pm 2,  \nonumber
\end{eqnarray}
\begin{equation}
\begin{split}
Q \approx & \frac{\ln l_0 - \ln l_-}{\ln l_+ - \ln l_-} \frac{S(l_+)-S(l_0)}{\ln l_+ - \ln l_0} + \\
& + \frac{\ln l_+ - \ln l_0}{\ln l_+ - \ln l_-} \frac{S(l_0)-S(l_-)}{\ln l_0 - \ln l_-},   \label{eq:Q}
\end{split}
\end{equation}
where $[\ ]$ means to round up the value. The $Q$'s for various $t_\downarrow$ and $L$ are computed and the results are presented in Fig.~\ref{fig:dmrg_BEv_Q}(a). For each curve, there exists a region where $Q$ quickly drops towards zero. This is the critical region where the phase transition occurs. For better illustration, we give the first derivatives in Fig.~\ref{fig:dmrg_BEv_Q}(b). For each line the peak locates the critical point $t_{\downarrow c}$. Then we perform scaling with $1/L$, as shown in the inset, and find that in the thermodynamic limit $t_{\downarrow c} \approx 0.32$.

The above studies show that the two-site entropy is not a sharp indicator for the phase transition of the half-filling AHM, though it shows distinct behaviors in the two phases. The block entropy, however, is useful for witnessing the phase transition.

\section{\label{sec:summary}Summary}

To summarize, we have studied the ground state phase diagram of the asymmetric Hubbard chain by means of entanglement. The middle-two-site entropy $S_2$ and the block entropy $S(l)$ are computed for open chains using the DMRG method.

In the case of away from half-filling, we found that $S_2$ is a good indicator of phase transition between the DW phase and PS phase. The $S_2$ function of $t_\downarrow$ displays a sharp change at the transition point. This phase transition is deduced to be the Landau type. We performed scaling studies and presented the phase diagram. Besides, The block entropy $S(l)$ is also a good indicator of phase transition as long as $l$ is long enough. The block entropy in the DW phase is higher than that in the PS phase.

On the other hand, in the case of half-filling, $S_2$ as a function of $t_\downarrow$ shows different behaviors in the Hubbard phase and FK phase. There is no sharp change of $S_2$ in any derivatives as the phase transition is Kosterlitz-Thouless like. Chains with different lengths were studied and we found that two families of curves are formed according to the parities of the chains due to singlet formations. Besides, the scaling behaviors of $S_2$ with chain length $L$ in the two phases are different. $S_2$ scales as $1/\sqrt{L}$ in the Hubbard phase but scales as $\exp(-L/\xi)$ in the FK phase. The block entropy $S(l)$ also shows different behaviors in the two phases. In the Hubbard phase $S(l)$ changes algebraically, while in the FK phase it changes exponentially. By applying the concepts in conformal field theory, we propose to compute the parameter $Q$ from the block entropies and that $Q$ is a valid indicator of the phase transition.

The above studies reflect certain correlations between entanglement and phase transition. They consolidate the idea that the entropy of a part of a system provides fruitful information about the whole system based on the superposition principle of quantum mechanics.

\begin{acknowledgments}
This work is supported by the Earmarked Grant for Research from the Research Grants Council of HKSAR, China (Projects No.~400906 and N\_CUHK204/05).
\end{acknowledgments}

\end{document}